\documentclass{Interspeech}

\interspeechcameraready 

\usepackage{subfigure} 
\usepackage{cite}
\usepackage{amsmath,amssymb,amsfonts}
\usepackage{algorithmic}
\usepackage{graphicx}
\usepackage{textcomp}
\usepackage{xcolor}
\usepackage[utf8]{inputenc}
\usepackage{float}
\usepackage{graphicx}
\usepackage{multirow}
\usepackage{xcolor}
\usepackage{hyperref}
\usepackage{acronym}
\usepackage{supertabular}
\usepackage{booktabs}
\usepackage{amssymb}
\usepackage{arydshln} 
\usepackage{color} 
\usepackage{amsmath} 
\usepackage{enumitem}

\title{Auto-Landmark: Acoustic Landmark Dataset and Open-Source Toolkit for Landmark Extraction}

\author[affiliation={1}]{Xiangyu}{Zhang}
\author[affiliation={1}]{Daijiao}{Liu}
\author[affiliation={3}]{Tianyi}{Xiao}
\author[affiliation={2}]{Cihan}{Xiao}
\author[affiliation={1}]{Tuende}{Szalay}
\author[affiliation={1}]{Mostafa}{Shahin}
\author[affiliation={1}]{Beena}{Ahmed}
\author[affiliation={1}]{Julien}{Epps}

\affiliation{}{The University of New South}{}
\affiliation{}{Johns Hopkins University}{}
\affiliation{}{Nanyang Technological University}{}
\email{xiangyu.zhang2@unsw.edu.au}

\usepackage{comment}

\begin{document}

\maketitle

\begin{abstract}
In the speech signal, acoustic landmarks identify times when the acoustic manifestations of the linguistically motivated distinctive features are most salient. Acoustic landmarks have been widely applied in various domains, including speech recognition, speech depression detection, clinical analysis of speech abnormalities, and the detection of disordered speech. However, there is currently no dataset available that provides precise timing information for landmarks, which has been proven to be crucial for downstream applications involving landmarks.  In this paper, we selected the most useful acoustic landmarks based on previous research and annotated the TIMIT dataset with them, based on a combination of phoneme boundary information and manual inspection. Moreover, previous landmark extraction tools were not open source or benchmarked, so to address this, we developed an open source Python-based landmark extraction tool and established a series of landmark detection baselines. The first of their kinds, the dataset with landmark precise timing information, landmark extraction tool and baselines are designed to support a wide variety of future research.
\end{abstract}

\section{Introduction}
\begin{figure}[t]
    \centering
    \includegraphics[width=0.48\textwidth]{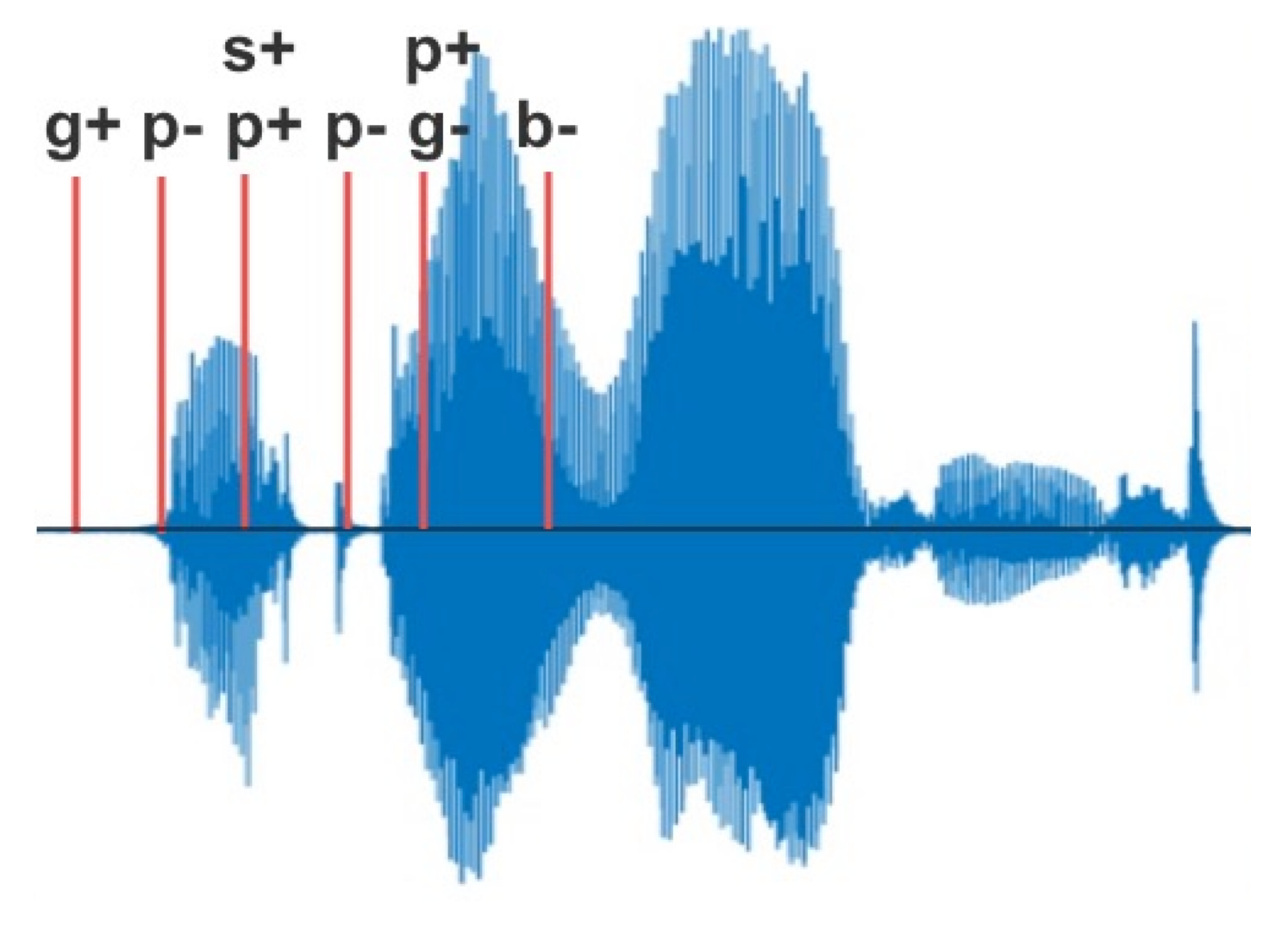}
    \caption{Example of acoustic landmarks: The speech signal is discretized into a series of tokens with  speech production information}

    \label{fig:landmark}
\end{figure}

In speech processing, frame-based methods are most commonly used to segment the speech waveform. This method treats each frame as the central unit of analysis, with a fixed set of speech attributes measured at each frame, has been used in many domain~\cite{li2025speech,li2025from}. However, because this approach relies on fixed-duration processing, it often overlooks important timing factors such as speaking rate and segmental duration. In contrast, acoustic landmark detection focuses on specific, acoustically significant points in the speech signal~\cite{liu1996landmark}, as illustrated in Figure~\ref{fig:landmark}. This method allows for the identification of acoustic landmarks somewhat independently of frames, while also providing valuable timing information for subsequent processing.

Acoustic landmarks have been proven highly effective across various fields. Initially applied in speech recognition~\cite{liu1996landmark,he2019ctc}, they have since been extended to the health domain, including applications such as depression detection~\cite{huang2019investigation,huang2018depression}, clinical analysis of speech abnormalities~\cite{ishikawa2023landmark}, and the detection of disordered speech~\cite{ishikawa2017toward}. Despite the success of acoustic landmarks in various fields, no dataset currently provides precise ground truth timing information for the occurrence of acoustic landmarks, making it difficult to produce standardized results on a baseline dataset~\cite{huang2019investigation,ishikawa2017toward}. To address this research gap, we selected the most useful acoustic landmarks based on previous research~\cite{zhang2024llms,huang2018depression,huang2019natural,ishikawa2023landmark,ishikawa2017toward} and annotated the TIMIT dataset~\cite{garofolo1993timit} with them, with some landmarks labeled according to phoneme boundary information and others manually annotated.

Moreover, previous landmark extraction tools were based on closed-source software, making it difficult to understand their inner workings, and had not been benchmarked. To overcome this limitation, we developed an open source Python-based acoustic landmark extraction tool - Auto-Landmark and established a series of acoustic landmark detection baselines to support future research.

\section{Related Work}

\subsection{Acoustic Landmarks}
The idea of acoustic landmarks originates from studies on distinctive features~\cite{garvin1953preliminaries}. Distinctive features offer a concise way to describe the sounds of a language at a subsegmental level. They are closely related to both acoustics and articulation, characterized by binary values. This minimal set of features allows for the differentiation of each segment within a language. Landmarks serve as indicators of the underlying segments, which group distinctive features into bundles. Landmarks are categorized into four groups: abrupt-consonantal, abrupt, non-abrupt, and vocalic. While scholars have slightly different definitions of acoustic landmarks,~\cite{boyce2012speechmark} expanded upon~\cite{liu1996landmark} by releasing a MATLAB-based landmark detection toolkit (SpeechMark), which has become the most widely used approach for landmark detection.

\subsection{Application of Acoustic Landmarks}

\renewcommand{\arraystretch}{1.5} 

\begin{table}[t]
\centering
\caption{Description of the five landmarks investigated.}
\vspace{4pt}
\begin{tabular}{l| p{0.65\linewidth}}
\hline
\hline

\textbf{Landmark} & \textbf{Description} \\
\hline
\multirow{2}{*}{g} & vibration of vocal folds start (+) or end (–) \\ 

\multirow{2}{*}{b} & onset (+) or offset (–) of existence of turbulent noise during obstruent regions \\

\multirow{1}{*}{s} & releases (+) or closures (–) of a nasal \\

\multirow{1}{*}{v} & voiced frication onset (+) or offset (–) \\ 

f & frication onset (+) or offset (–) \\
\hline 

\end{tabular}
\label{landmark_table}
\end{table}
Acoustic landmarks were initially used in speech recognition~\cite{liu1996landmark, he2019ctc}, with some researchers exploring the combination of acoustic landmarks with CTC in this context~\cite{he2019ctc}. Following their success in speech recognition, acoustic landmarks have been applied to various other domains. Acoustic landmarks have been widely applied in various health-related studies. For example,~\cite{huang2019investigation, huang2018depression} utilized acoustic landmark timing information for depression detection.~\cite{zhang2024llms,zhang2025pre,zhang2025speecht} further extended the use of acoustic landmarks by tokenizing speech for large language model-based systems in depression detection.~\cite{ishikawa2017toward} used acoustic landmark time duration to describe abnormalities in speech production that affect a speaker’s intelligibility.~\cite{ishikawa2023landmark} evaluated the feasibility of differentiating conversational and clear speech produced by individuals with muscle tension dysphonia (MTD) using landmark-based analysis of speech.

\section{Acoustic Landmark Dataset}
\subsection{Dataset and Landmark Selection}
As shown in Table~\ref{landmark_table}, we selected five different types of acoustic landmarks based on previous research~\cite{zhang2024llms, huang2019investigation, huang2019natural, huang2018depression,ishikawa2023landmark,ishikawa2017toward}, each with the onset and offset states. These landmarks are g(lottis), s(sonorant), f(ricative), v(voiced fricative), and b(ursts), which are pointing in time where different abrupt articulatory events occur.

The \textbf{g (glottis)} landmark captures the start or end of the vocal fold vibration, which is crucial in identifying voiced phonemes where the vocal cords are active. The \textbf{b (bursts)} landmark marks the onset or offset of turbulent noise during obstruent regions, such as plosives or stops, which are significant for identifying consonant sounds. The \textbf{s (sonorant)} landmark indicates the releases or closures of nasal sounds and glides [l], which are produced with a continuous airflow through the vocal tract. Proper identification of these sonorants is important for understanding the nasal quality in speech. The \textbf{v (voiced fricative)} landmark captures the onset or offset of voiced frication, with voiced fricatives like [v] and [z] involving continuous vibration of the vocal cords while producing a turbulent airflow. Lastly, the \textbf{f (fricative)} landmark indicates the onset or offset of frication, with fricatives like [f] and [s] characterized by the turbulent airflow created by a narrow constriction in the vocal tract~\cite{nathan1998sounds,culbertson2024fundamentals}.

\subsection{Labeling Convention}
As discussed in the previous section, landmarks are closely associated with phonemes, which makes the TIMIT dataset~\cite{garofolo1993timit} with its manually annotated phoneme boundaries an ideal choice for annotation. We utilized Praat~\cite{boersma2001speak} to facilitate the labeling process. We first identified the boundaries of the phonemes associated with landmark, then used Praat to help determine the exact timing of these landmarks.

\begin{table}[t]
\centering
\caption{Rules for Annotating Data}
\vspace{4pt}
\begin{tabular}{l| p{0.65\linewidth}}
\hline
\hline

\textbf{Landmark} & \textbf{Details} \\
\hline
v+ & Start time of voiced frication phone \\
v- & End time of voiced frication phone \\
f+ & Start time of frication phone \\
f- & End time of frication phone \\
s+ & Start time of nasal or [l] phone \\
s- & End time of nasal or [l] phone \\
b+ & Sudden energy increase in stop and affricate in the spectrum \\
b- & Sudden energy decrease in stop and affricate in the spectrum \\
g+ & Start time of voiced phones (vowels, voiced fricatives, nasals, b, d, g) \\
g- & End of glottal phone, or change to voiceless stops, fricatives, or affricates \\
\hline

\end{tabular}
\label{annotation_rules_table}
\end{table}

Table~\ref{annotation_rules_table} outlines the rules we followed for annotating acoustic landmarks. Each landmark corresponds to specific phonetic events and transitions within the speech signal. The v+ and v- landmarks indicate the start and end times of voiced fricatives, such as [z], [zh], [v], and [dh]. Similarly, f+ and f- mark the start and end times of fricatives like [s], [sh], [f], and [th]. For nasals and the glides [l], the s+ and s- landmarks represent the start and end times, covering phonemes like [m], [n], [ng], [em], [en], [eng], and [nx]. The b+ and b- landmarks capture the sudden energy changes in stops and affricates, including [b], [d], [g], [p], [t], [k], [dx], [q], [jh], and [ch] and manually labeled. Finally, the g+ and g- landmarks denote the start and end times of voiced phones, encompassing a range of vowels (e.g., [iy], [ih], [eh], [ey], [ae], [aa], [aw], [ay], [ah], [ao], [oy], [ow], [uh], [uw], [ux], [er], [ax], [ix], [axr], [ax-h]), voiced fricatives, nasals, and stops such as [b], [d], and [g]. The g- landmark also indicates the transition from voiced to voiceless sounds.

\subsection{Statistics of Landmark Dataset}
\begin{figure}[t!]
    \centering
    \subfigcapskip=0pt  
    \subfigure{%
        \includegraphics[width=0.48\columnwidth]{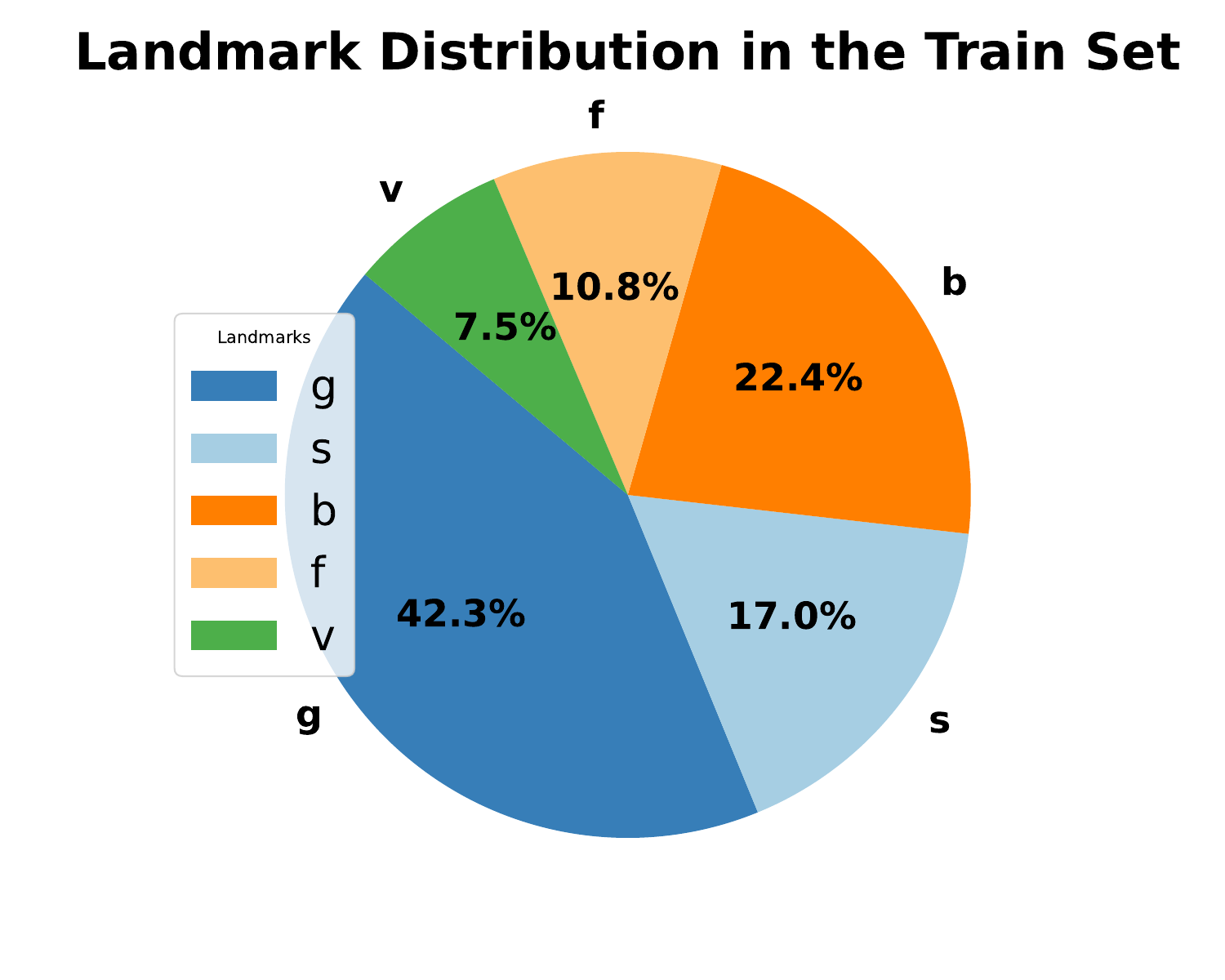}
        \label{fig:train_pie_chart}
    }
    \hfill
    \subfigure{%
        \includegraphics[width=0.48\columnwidth]{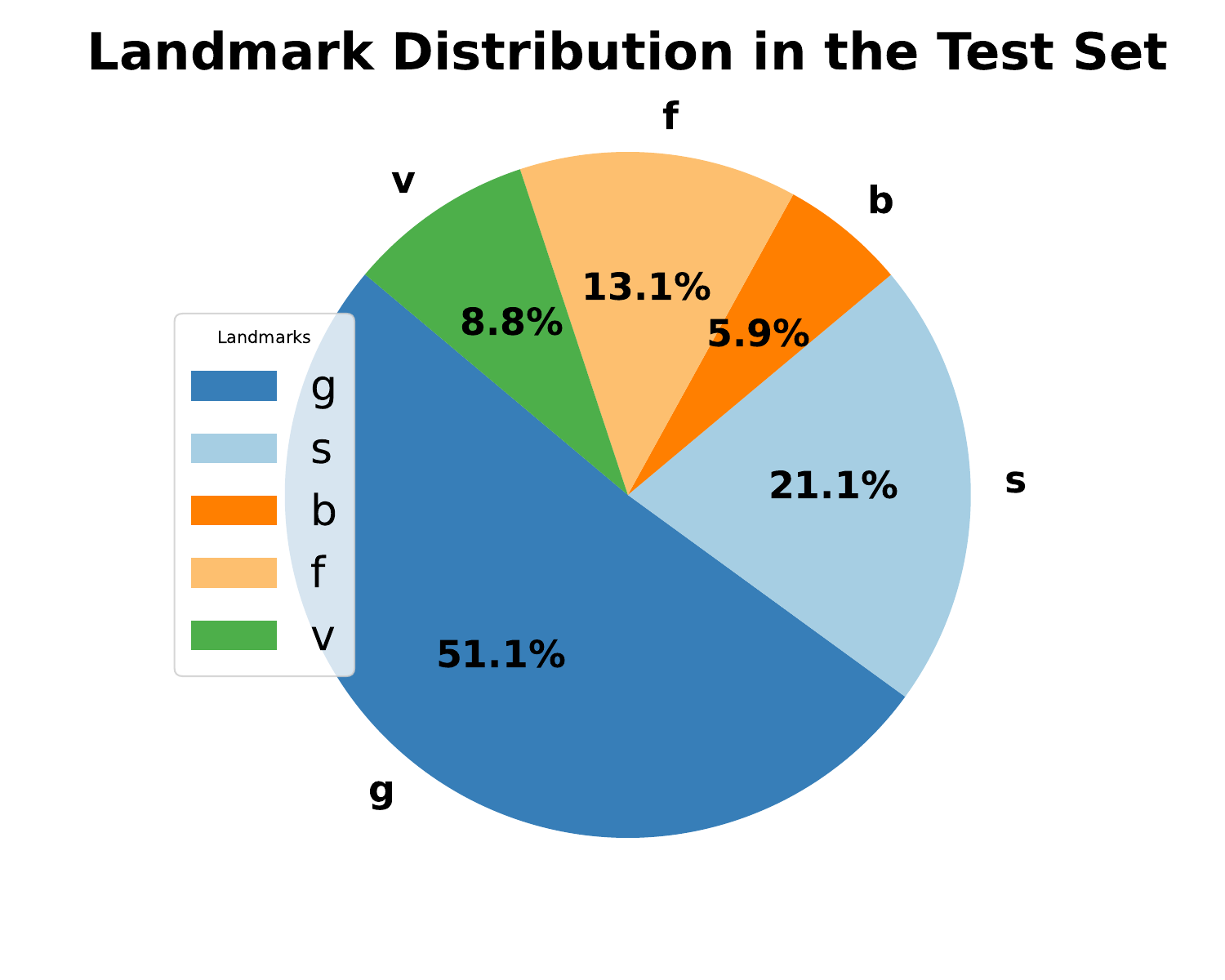}
        \label{fig:test_pie_chart}
    }
    \caption{Landmark distribution in the TIMIT dataset~\cite{garofolo1993timit} for the training and test sets. Right is Landmark distribution in the training set, left is Landmark distribution in the test set}
    \label{fig:landmark_distribution}
\end{figure}

Figure~\ref{fig:landmark_distribution} illustrates the proportion of each landmark in the dataset. As shown, the g landmark has the highest proportion in both the training and test sets. However, the proportion of the b landmark in the test set is significantly lower than in the training set. We found that this discrepancy is due to the lower proportion of phonemes associated with the b landmark in the test set compared with the train set.

\section{Landmark Detection System}
We provide both signal processing and deep learning methods for landmark detection. For the former, we developed an open-source Python tool to address the gap in fully Python based open-source solutions. For the latter, we advanced the field by incorporating landmark detection into cutting-edge research and provided an ESPnet~\cite{watanabe2018espnet} recipe to facilitate this.

\subsection{Filter-based Method}
Our implementation of the signal processing-based method is based on~\cite{zhang2024llms,liu1996landmark,boyce2012speechmark}. Initially, the spectrogram is divided into six frequency bands. Landmarks are identified through energy changes within these bands using a two-pass strategy. Different landmarks are determined by analyzing either a single band or a combination of multiple bands.
 
The \textbf{g landmark} is identified when both the coarse and fine filters exhibit a peak in band 1. The \textbf{b landmark}, in an unvoiced segment (not between +g and the next -g), it is identified if at least three out of five frequency bands show simultaneous power increases of no less than 6 dB in both coarse and fine filters. The \textbf{s landmark} is identified in a voiced segment (between +g and the next -g) under the same conditions: simultaneous power increases of no less than 6 dB in at least three out of five frequency bands in both coarse and fine filters. The \textbf{f+} and \textbf{v+ landmarks} are detected by identifying a 6 dB power increase in at least three high-frequency bands (4, 5, 6) and a power decrease in low-frequency bands (2, 3). Conversely, for \textbf{f-} and \textbf{v-}, the criteria involve a 6 dB power decrease in the high-frequency bands and a power increase in the low-frequency bands. The key distinction is that frication landmarks are detected in unvoiced segments (b landmark), while voiced frication landmarks are detected in voiced segments (s landmark). The details of the method can be found in~\cite{zhang2024llms,liu1996landmark,boyce2012speechmark}.

\subsection{Deep Learning Based Method}
To date, landmark detection has not been performed with deep learning models. Our first consideration is how to model this problem. In this paper, since both landmark detection and speech recognition tasks involve converting speech signals into symbols, we approach landmark detection as a speech recognition task. We employ a Hybrid CTC/attention end-to-end architecture~\cite{watanabe2017hybrid}. We utilized two types of models as the Encoder: the Conformer~\cite{gulati2020conformer} and ConExtBimamba~\cite{zhang2024mamb,zhang2025rethinking,chen2025selective}. For the decoder, we employed the Transformer model~\cite{vaswani2017attention}. Given the success of self-supervised models in various speech tasks, we therefore explored how they perform on the landmark detection task. Specifically, we utilized the wav2vec2.0 large model~\cite{baevski2020wav2vec}. We followed the Superb setup~\cite{yang21c_interspeech}, using the weighted sum feature from different layers of wav2vec2.0.

\begin{figure}[t]
    \centering
    \includegraphics[width=0.5\textwidth]{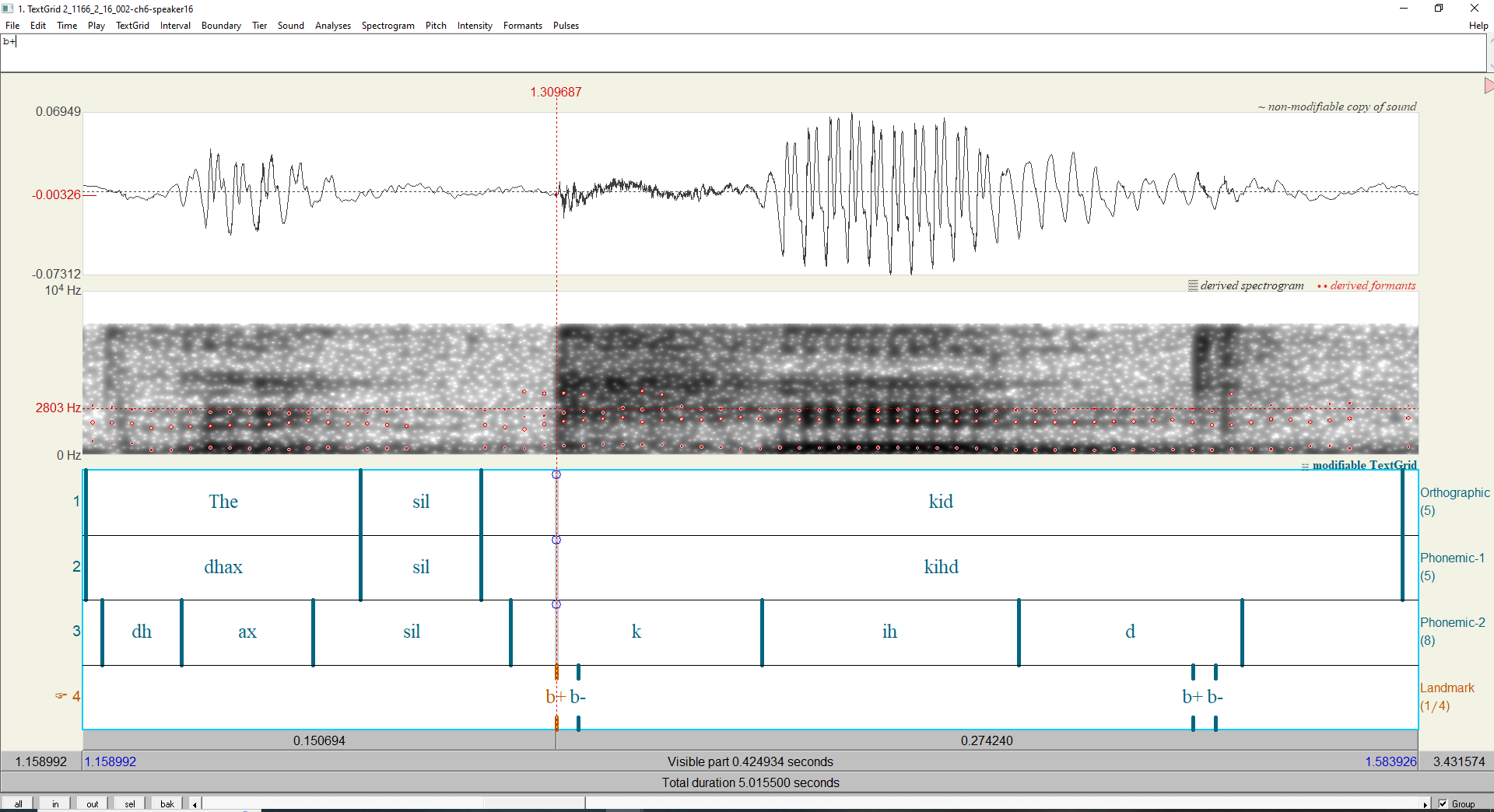}
    \caption{Example of labeling Using Praat: the b landmark is marked when the energy suddenly changes}

    \label{fig:software}
\end{figure}

\subsection{Toolkit for Landmark Extraction}

Previous landmark extraction tools were not open source, making it difficult to understand their inner workings. To address this, we developed an open-source Python-based tool, \textbf{Auto-Landmark}, for landmark extraction.

\noindent \textit{C.1) Directory Structure}

\noindent Auto-Landmark adopts a structure similar to the directory organization of ESPnet~\cite{watanabe2018espnet}. There are four main folders: \textbf{egs}, which is organized Recipe for each dataset; \textbf{methods}, which stores different landmark extraction methods; \textbf{utils}, containing commonly used functions; and \textbf{visualisation}, which houses functions related to visualization. There are two datasets now supported by Auto-landmark, TIMIT~\cite{garofolo1993timit} and DAIC-WOZ~\cite{devault2014simsensei}

\noindent \textit{C.2) Toolkit workflow}

\noindent\textbf{Stage 1-2 Data preparation:} The initial data format follows the Kaldi style. After preprocessing, all speech data is converted into a standard format for resampling and segmentation.

\noindent \textbf{Stage 3 Landmark Extraction:} At this stage, the primary task is to execute different algorithms for landmark extraction. So far, we have provided two distinct types of landmarks: basic and advanced. The main difference between the two lies in the implementation of the filters and the smoothing methods. The \textbf{basic method} applies moving window average smoothing, a simpler approach that focuses on noise reduction in energy signals. This method is effective for peak detection but offers less flexibility and control, making it more suitable for straightforward tasks. On the other hand, the \textbf{advanced method} utilizes linear convolution smoothing with customizable kernels, such as Hanning windows, providing greater control and adaptability for handling more complex data. For more details on their differences, please refer to the code.

\noindent \textbf{Optional Visualization:} We have provided additional visualization tools to help health researchers in analyzing cases.

\section{Results and Discussion}
\subsection{Experimental Setup}

\noindent \textbf{Signal Processing Method}: Our two-pass filter utilizes six band energies, specifically in the ranges of 0.0–0.4, 0.8–1.5, 1.2–2.0, 2.0–3.5, 3.5–5.0, and 5.0–8.0 kHz. Following previous work~\cite{zhang2024llms, liu1996landmark, boyce2012speechmark}, the coarse processing uses 20ms smoothing with an 8 dB threshold for peak detection. In fine processing, 10ms smoothing is applied, and peak detection targets thresholds within the range of 5 dB to 8 dB.

\noindent \textbf{Deep Learning Method}: We conducted our experiments using a V100 32G GPU, with the model configuration based on the best results from the AN4 dataset~\cite{acero1990environmental} in ESPnet, given that both TIMIT and AN4 are small datasets.  The training utilized a single V100 32G GPU with batch bins of 1,000,000, using the Adam optimizer with a learning rate of 0.001 and a warm-up scheduler for 2500 steps. The Conformer/ConExtBimamba encoder was configured with an output size of 256, 2048 linear units, and 6 blocks. The Transformer decoder had 4 attention heads, 2048 linear units, 6 blocks, and a dropout rate of 0.1.
\begin{table}[t!]
\centering
\caption{Landmark detection results for different methods}
\vspace{8pt}
\begin{tabular}{lccc}
\hline
\hline
\textbf{Method} &  & \textbf{LER}(\textdownarrow) \\
\midrule
Singal Processing Method & &  \\
\quad $-$ SpeechMark~\cite{boyce2012speechmark} & & 56.53 \\
\quad $-$ Auto-Landmark  &  & 56.73 &\\
\hdashline
Deep Learning Method &  &   \\
\quad $-$ Conformer~\cite{gulati2020conformer} & & 33.0 \\
\quad $-$ ConBiMamba~\cite{zhang2024mamb} & & \textbf{31.3}\\
\quad $-$ Wav2vec2 $+$ Conformer & & 48.3\\
\quad $-$ Wav2vec2 $+$ ConBiMamba & & 44.6\\
\toprule[1.5pt]
\end{tabular}
\vspace{-16pt}
\label{Main Result}
\end{table}

\subsection{Baseline Result}
Table~\ref{Main Result} compares the landmark detection results from the two methods, evaluated by Landmark Error Rate (LER). LER was calculated based on patterns without considering landmark timing. Additionally, following previous conventions, we did not differentiate between onset (+) and offset (–) landmarks~\cite{liu1996landmark}.All the deep learning models produced lower LER compared to both signal processing methods, with the ConExtBiMamba model outperforming the others. Typically, the overall performance of models improves when using self-supervised models. However, as shown in Table~\ref{Main Result}, the performance of both the Conformer and ConExtBimamba models decreased when combined with wav2vec2, contrary to expectations. This issue arose because several landmarks occur in close temporal proximity to each other, even simultaneously. This means that numerous landmarks can appear within a 10ms window. However, as wav2vec uses a 20ms window with a 10ms stride, it potential misses some landmarks.

\subsection{Insights from Self-supervised Weights}
\begin{figure}[t]
    \centering
    \includegraphics[width=0.45\textwidth]{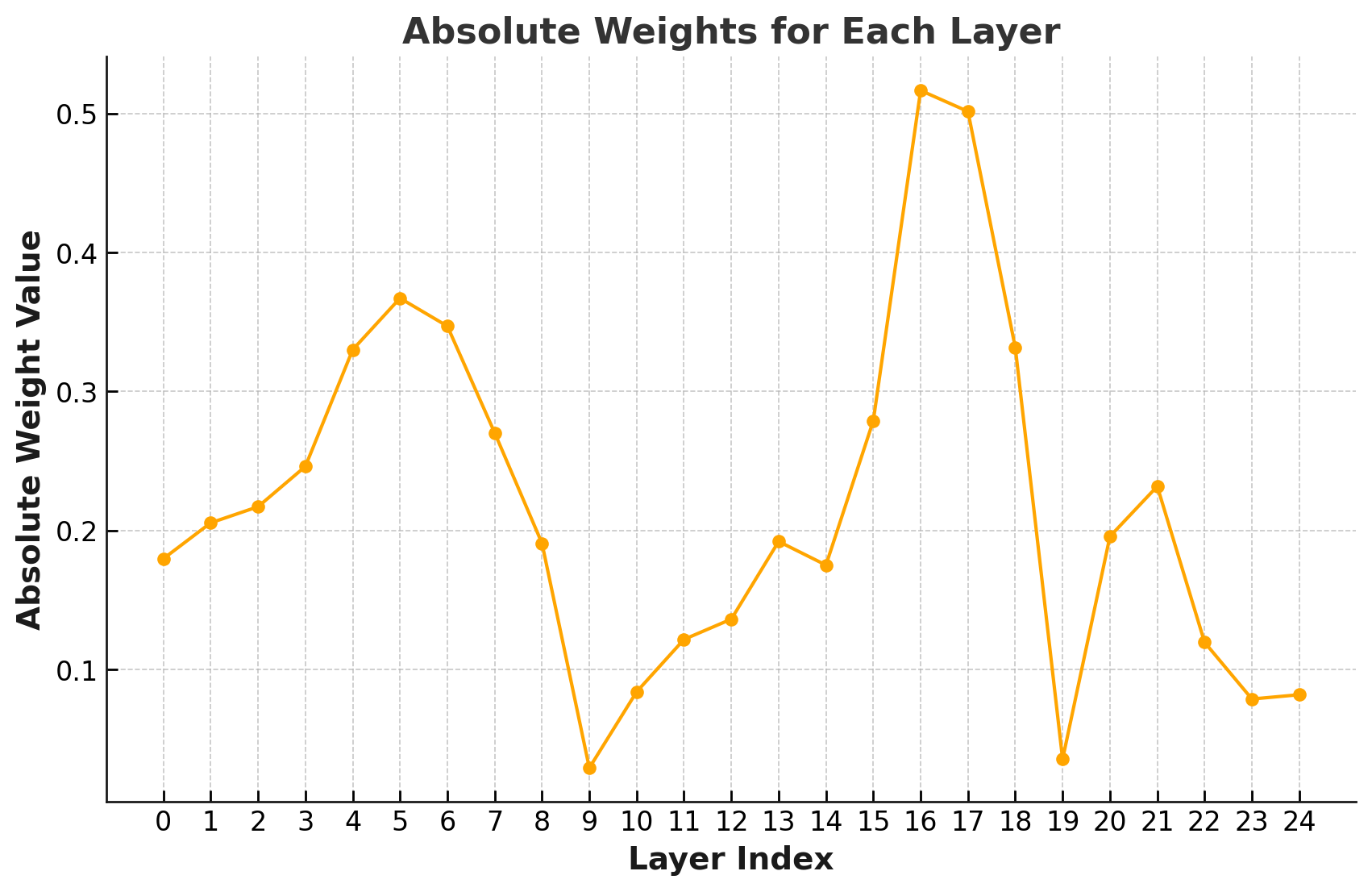}
    \caption{Layer Weight of wav2vec2.0 for Landmark Detection}

    \label{layer weight}
\end{figure}

Previous research has shown that when using the weighted sum feature of self-supervised models for downstream tasks, the weights can be used to analyze the properties of the models and tasks~\cite{li2023quantitative}. Figure~\ref{layer weight} shows the feature weights from different layers when using wav2vec2 large combined with the Conformer for landmark detection. Layers 16, 17, and 18 have the highest weights. Previous studies using CCA and mutual information to analyze wav2vec2 found that phonetic information is most salient around layer 18~\cite{pasad2021layer}. This suggests that wav2vec2 treats landmark detection as a task highly related to phonetic information. However, as described by the definition of landmarks, they primarily denote points of abrupt change in the speech signal. Thus, low-level features in speech are also crucial. Due to the limitations of wav2vec's window size, many consecutive abrupt changes are not captured. From Figure~\ref{layer weight}, we can see that, unlike the layers associated with phonetic information identified by CCA analysis, the weights of shallow features remain significant for landmark detection. According to~\cite{pasad2023comparative}, the shallow features of wav2vec are more similar to low-level features, indicating that the model captures effects that are not solely related to phonetic information.

\section{Conclusion}

In this study, we made several key contributions to the field of acoustic landmark detection. First, we provided the first publicly available dataset with manually annotated acoustic landmarks and precise timing ground truth. Second, we also developed the first Python-based open-source implementation of the landmark detection methods originally proposed by ~\cite{liu1996landmark}. Finally, we established new baselines for landmark detection with deep learning models and comprehensively benchmarked these methods. Our analysis demonstrated that effective landmark detection requires both phonetic and low-level feature information. This dataset and the accompanying tools will significantly benefit research into conditions that affect speech timing, such as dementia, Parkinson's, intoxication, speech impairment, and language learning. Previous research has shown that timing information is crucial for the application of landmarks in mental health~\cite{huang2019investigation, huang2019natural}, making our timing-based benchmarking particularly valuable. Moreover, since CTC-based models struggle to capture timing information, developing deep learning models that can more precisely estimate the timing of landmarks will be an important direction for future research.

\section{Acknowledgments}
This work was supported by the Australian Research Council Discovery Project DP230101184.

\bibliographystyle{IEEEtran}
\bibliography{mybib}

\begin{thebibliography}{10}
\providecommand{\url}[1]{#1}
\csname url@samestyle\endcsname
\providecommand{\newblock}{\relax}
\providecommand{\bibinfo}[2]{#2}
\providecommand{\BIBentrySTDinterwordspacing}{\spaceskip=0pt\relax}
\providecommand{\BIBentryALTinterwordstretchfactor}{4}
\providecommand{\BIBentryALTinterwordspacing}{\spaceskip=\fontdimen2\font plus
\BIBentryALTinterwordstretchfactor\fontdimen3\font minus \fontdimen4\font\relax}
\providecommand{\BIBforeignlanguage}[2]{{%
\expandafter\ifx\csname l@#1\endcsname\relax
\typeout{** WARNING: IEEEtran.bst: No hyphenation pattern has been}%
\typeout{** loaded for the language `#1'. Using the pattern for}%
\typeout{** the default language instead.}%
\else
\language=\csname l@#1\endcsname
\fi
#2}}
\providecommand{\BIBdecl}{\relax}
\BIBdecl

\bibitem{li2025speech}
H.~Li, J.~Q. Yip, T.~Fan, and E.~S. Chng, ``Speech enhancement using continuous embeddings of neural audio codec,'' in \emph{ICASSP 2025-2025 IEEE International Conference on Acoustics, Speech and Signal Processing (ICASSP)}.\hskip 1em plus 0.5em minus 0.4em\relax IEEE, 2025, pp. 1--5.

\bibitem{li2025from}
H.~Li, Y.~Hu, C.~Chen, S.~M. Siniscalchi, S.~Liu, and E.~S. Chng, ``From kan to gr-kan: Advancing speech enhancement with kan-based methodology,'' \emph{arXiv preprint arXiv:2412.17778}, 2025.

\bibitem{liu1996landmark}
S.~A. Liu, ``Landmark detection for distinctive feature-based speech recognition,'' \emph{The Journal of the Acoustical Society of America}, vol. 100, no.~5, pp. 3417--3430, 1996.

\bibitem{he2019ctc}
D.~He, X.~Yang, B.~P. Lim, Y.~Liang, M.~Hasegawa-Johnson, and D.~Chen, ``When ctc training meets acoustic landmarks,'' in \emph{ICASSP 2019-2019 IEEE International Conference on Acoustics, Speech and Signal Processing (ICASSP)}.\hskip 1em plus 0.5em minus 0.4em\relax IEEE, 2019, pp. 5996--6000.

\bibitem{huang2019investigation}
Z.~Huang, J.~Epps, and D.~Joachim, ``Investigation of speech landmark patterns for depression detection,'' \emph{IEEE transactions on affective computing}, vol.~13, no.~2, pp. 666--679, 2019.

\bibitem{huang2018depression}
Z.~Huang, J.~Epps, D.~Joachim, and M.~Chen, ``Depression detection from short utterances via diverse smartphones in natural environmental conditions.'' in \emph{INTERSPEECH}, 2018, pp. 3393--3397.

\bibitem{ishikawa2023landmark}
K.~Ishikawa, M.~Pietrowicz, S.~Charney, and D.~Orbelo, ``Landmark-based analysis of speech differentiates conversational from clear speech in speakers with muscle tension dysphonia,'' \emph{JASA Express Letters}, vol.~3, no.~5, 2023.

\bibitem{ishikawa2017toward}
K.~Ishikawa, J.~MacAuslan, and S.~Boyce, ``Toward clinical application of landmark-based speech analysis: Landmark expression in normal adult speech,'' \emph{The Journal of the Acoustical Society of America}, vol. 142, no.~5, pp. EL441--EL447, 2017.

\bibitem{zhang2024llms}
X.~Zhang, H.~Liu, K.~Xu, Q.~Zhang, D.~Liu, B.~Ahmed, and J.~Epps, ``When llms meets acoustic landmarks: An efficient approach to integrate speech into large language models for depression detection,'' \emph{arXiv preprint arXiv:2402.13276}, 2024.

\bibitem{huang2019natural}
Z.~Huang, J.~Epps, D.~Joachim, and V.~Sethu, ``Natural language processing methods for acoustic and landmark event-based features in speech-based depression detection,'' \emph{IEEE Journal of selected topics in Signal Processing}, vol.~14, no.~2, pp. 435--448, 2019.

\bibitem{garofolo1993timit}
J.~S. Garofolo, ``Timit acoustic phonetic continuous speech corpus,'' \emph{Linguistic Data Consortium, 1993}, 1993.

\bibitem{garvin1953preliminaries}
P.~L. Garvin, ``Preliminaries to speech analysis: The distinctive features and their correlates,'' 1953.

\bibitem{boyce2012speechmark}
S.~Boyce, H.~J. Fell, and J.~MacAuslan, ``Speechmark: Landmark detection tool for speech analysis.'' in \emph{INTERSPEECH}, 2012, pp. 1894--1897.

\bibitem{zhang2025pre}
X.~Zhang, B.~Ahmed, and J.~Epps, ``Why pre-trained models fail: Feature entanglement in multi-modal depression detection,'' \emph{arXiv preprint arXiv:2503.06620}, 2025.

\bibitem{zhang2025speecht}
X.~Zhang, H.~Liu, Q.~Zhang, B.~Ahmed, and J.~Epps, ``Speecht-rag: Reliable depression detection in llms with retrieval-augmented generation using speech timing information,'' \emph{arXiv preprint arXiv:2502.10950}, 2025.

\bibitem{nathan1998sounds}
G.~S. Nathan, ``The sounds of the world's languages,'' 1998.

\bibitem{culbertson2024fundamentals}
W.~Culbertson, \emph{Fundamentals of the Speech and Language Sciences}.\hskip 1em plus 0.5em minus 0.4em\relax Routledge, 2024.

\bibitem{boersma2001speak}
P.~Boersma and V.~Van~Heuven, ``Speak and unspeak with praat,'' \emph{Glot International}, vol.~5, no. 9/10, pp. 341--347, 2001.

\bibitem{watanabe2018espnet}
\BIBentryALTinterwordspacing
S.~Watanabe, T.~Hori, S.~Karita, T.~Hayashi, J.~Nishitoba, Y.~Unno, N.~{Enrique Yalta Soplin}, J.~Heymann, M.~Wiesner, N.~Chen, A.~Renduchintala, and T.~Ochiai, ``{ESPnet}: End-to-end speech processing toolkit,'' in \emph{Proceedings of Interspeech}, 2018, pp. 2207--2211. [Online]. Available: \url{http://dx.doi.org/10.21437/Interspeech.2018-1456}
\BIBentrySTDinterwordspacing

\bibitem{watanabe2017hybrid}
S.~Watanabe, T.~Hori, S.~Kim, J.~R. Hershey, and T.~Hayashi, ``Hybrid ctc/attention architecture for end-to-end speech recognition,'' \emph{IEEE Journal of Selected Topics in Signal Processing}, vol.~11, no.~8, pp. 1240--1253, 2017.

\bibitem{gulati2020conformer}
A.~Gulati, J.~Qin, C.-C. Chiu, N.~Parmar, Y.~Zhang, J.~Yu, W.~Han, S.~Wang, Z.~Zhang, Y.~Wu \emph{et~al.}, ``Conformer: Convolution-augmented transformer for speech recognition,'' \emph{arXiv preprint arXiv:2005.08100}, 2020.

\bibitem{zhang2024mamb}
X.~Zhang, Q.~Zhang, H.~Liu, T.~Xiao, X.~Qian, B.~Ahmed, E.~Ambikairajah, H.~Li, and J.~Epps, ``Mamba in speech: Towards an alternative to self-attention,'' \emph{IEEE Transactions on Audio, Speech and Language Processing}, 2025.

\bibitem{zhang2025rethinking}
X.~Zhang, J.~Ma, M.~Shahin, B.~Ahmed, and J.~Epps, ``Rethinking mamba in speech processing by self-supervised models,'' in \emph{ICASSP 2025-2025 IEEE International Conference on Acoustics, Speech and Signal Processing (ICASSP)}.\hskip 1em plus 0.5em minus 0.4em\relax IEEE, 2025, pp. 1--5.

\bibitem{chen2025selective}
M.~Chen, Q.~Zhang, M.~Wang, X.~Zhang, H.~Liu, E.~Ambikairaiah, and D.~Chen, ``Selective state space model for monaural speech enhancement,'' \emph{IEEE Transactions on Consumer Electronics}, 2025.

\bibitem{vaswani2017attention}
A.~Vaswani, N.~Shazeer, N.~Parmar, J.~Uszkoreit, L.~Jones, A.~N. Gomez, {\L}.~Kaiser, and I.~Polosukhin, ``Attention is all you need,'' \emph{Advances in neural information processing systems}, vol.~30, 2017.

\bibitem{baevski2020wav2vec}
A.~Baevski, Y.~Zhou, A.~Mohamed, and M.~Auli, ``wav2vec 2.0: A framework for self-supervised learning of speech representations,'' \emph{Advances in neural information processing systems}, vol.~33, pp. 12\,449--12\,460, 2020.

\bibitem{yang21c_interspeech}
S.~wen Yang, P.-H. Chi, Y.-S. Chuang, C.-I.~J. Lai, K.~Lakhotia, Y.~Y. Lin, A.~T. Liu, J.~Shi, X.~Chang, G.-T. Lin, T.-H. Huang, W.-C. Tseng, K.~tik Lee, D.-R. Liu, Z.~Huang, S.~Dong, S.-W. Li, S.~Watanabe, A.~Mohamed, and H.~yi~Lee, ``{SUPERB: Speech Processing Universal PERformance Benchmark},'' in \emph{Proc. Interspeech 2021}, 2021, pp. 1194--1198.

\bibitem{devault2014simsensei}
D.~DeVault, R.~Artstein, G.~Benn, T.~Dey, E.~Fast, A.~Gainer, K.~Georgila, J.~Gratch, A.~Hartholt, M.~Lhommet \emph{et~al.}, ``Simsensei kiosk: A virtual human interviewer for healthcare decision support,'' in \emph{Proceedings of the 2014 international conference on Autonomous agents and multi-agent systems}, 2014, pp. 1061--1068.

\bibitem{acero1990environmental}
A.~Acero and R.~M. Stern, ``Environmental robustness in automatic speech recognition,'' in \emph{International Conference on Acoustics, Speech, and Signal Processing}.\hskip 1em plus 0.5em minus 0.4em\relax IEEE, 1990, pp. 849--852.

\bibitem{li2023quantitative}
S.~S. Li, B.~Xu, X.~Zhang, H.~Liu, W.~Chao, and P.~Garcia, ``A quantitative approach to understand self-supervised models as cross-lingual feature extracters,'' in \emph{Proceedings of the 6th International Conference on Natural Language and Speech Processing (ICNLSP 2023)}, 2023, pp. 200--211.

\bibitem{pasad2021layer}
A.~Pasad, J.-C. Chou, and K.~Livescu, ``Layer-wise analysis of a self-supervised speech representation model,'' in \emph{2021 IEEE Automatic Speech Recognition and Understanding Workshop (ASRU)}.\hskip 1em plus 0.5em minus 0.4em\relax IEEE, 2021, pp. 914--921.

\bibitem{pasad2023comparative}
A.~Pasad, B.~Shi, and K.~Livescu, ``Comparative layer-wise analysis of self-supervised speech models,'' in \emph{ICASSP 2023-2023 IEEE International Conference on Acoustics, Speech and Signal Processing (ICASSP)}.\hskip 1em plus 0.5em minus 0.4em\relax IEEE, 2023, pp. 1--5.

\end{thebibliography}

\end{document}